\documentclass[10pt, conference, compsocconf]{IEEEtran}

\usepackage{graphicx}
\usepackage{multirow}
\usepackage{array}
\usepackage{subfigure}
\usepackage{algorithm}
\usepackage{algorithmic}
\usepackage{graphicx}
\usepackage{amsmath}
\usepackage{subfigure}
\usepackage{rotating}
\usepackage{multirow}
\usepackage{array}
\usepackage{rotating}
\usepackage{auto-pst-pdf}

\begin{document}
\title{HSEP: Heterogeneity-aware Hierarchical\\ Stable Election Protocol for WSNs}

\author{\IEEEauthorblockN{A. A. Khan, N. Javaid, U. Qasim$^{\ddag}$, Z. Lu$^{\$}$, Z. A. Khan$^{\$}$\\}

        COMSATS Institute of Information Technology, Islamabad, Pakistan. \\
        $^{\ddag}$University of Alberta, Alberta, Canada.\\
        $^{\$}$Faculty of Engineering, Dalhousie University, Halifax, Canada
        }

\maketitle
\begin{abstract}
Wireless Sensor Networks (WSNs) are increasing to handle complex situations and functions. In these networks some of the nodes become Cluster Heads (CHs) which are responsible to aggregate data of from cluster members and transmit it to Base Stations (BS). Those clustering techniques which are designed for homogenous network are not enough efficient for consuming energy. Stable Election Protocol (SEP) introduces heterogeneity in WSNs, consisting of two type of nodes. SEP is based on weighted election probabilities of each node to become CH according to remaining energy of nodes. We propose Heterogeneity-aware Hierarchal Stable Election Protocol (HSEP) having two level of energies. Simulation results show that HSEP prolongs stability period and network lifetime, as compared to conventional routing protocols and having higher average throughput than selected clustering protocols in WSNs.
\end{abstract}


\section{Introduction}

WSNs are being continuously used in new applications in various areas, like, remote and hostile regions. In military, these networks are used for battle field surveillance, monitoring enemy territory and detection of attacks. Another applications of WSNs is in health sectors where patient wear small sensors on body for physiological data. To maintain reliable information delivery, WSNs require some efficient routing and MAC protocols. Two tasks: Data Aggregation and Information Fusion which are necessary for efficient and effective communication between sensor nodes are carried-out by CH within a cluster. Only processed and concise information is delivered to BS to reduce communication energy and prolong lifetime of network with optimal data delivery.

In Direct Transmission (DT) sensor nodes transmit directly to BS. As a result, nodes which are faraway from BS die first because they lose more energy in transmitting data to BS due to large distance between nodes and BS. Another technique is Minimum Transmission Energy (MTE), in which data is routed over minimum cost routes where minimum transmission energy is extended. Using MTE, nodes near to BS act as relay having higher probability to die first than other nodes which are far from BS. An inefficient use of available energy leads to poor performance and short life cycle of network, therefore, energy in these sensors is an important resource and must be managed in an efficient manner. Another routing protocol discussed by authors is Low-Energy Adaptive Clustering Hierarchy (LEACH) which is used in homogeneous networks. However, it uses clustering technique which is not used by DT or MTE. CHs are elected probabilistically in LEACH where each node becomes a CH according to a random number when compare with defined threshold.

We propose a new protocol, HSEP, which reduces transmission cost from CH to BS. The proposed protocol is heterogeneous-aware in the sense that election probabilities are weighted by the initial energy of a node relative to that of other nodes in the network. This enhance time interval  before death of first node is refereed as stability period. This period is very important for many applications where reliable feed-back from the sensor network is necessary. HSEP minimizes transmission energy by choosing secondary CHs from existing primary CHs in each round and these secondary CH are elected on bases of some probability.

\section{Related Work and Motivation}
In \cite{1}, first clustering based routing protocol for WSNs is presented: LEACH. This protocol uses random rotation of CH to evenly distribute energy load between sensor nodes to enhance stability period and network lifetime. LEACH is designed for homogeneous networks; here the term homogeneity refers to the nodes having same initial energy. Due to lack of heterogeneity in LEACH, authors do not discuss hierarchal clustering.

Another clustering based protocol known as SEP is proposed in [2]. In this clustering technique, to evenly balance load between nodes and heterogeneity in terms of their energy issues are tackled. It uses two types of nodes: advance nodes and normal nodes. Advance nodes have more chances to become a CH than normal nodes. This technique prolongs network life through increasing stability  period. However, we discuss hierarchal clustering technique to minimize transmission distance among CHs and BS which is not adopted in SEP.

WSNs require minimum energy consumption to prolong network life and stability period. Authors in \cite{3}, use clustering base routing protocol with three level of node heterogeneity in terms of energy to prolong network lifetime and stability period. However, we discuss hierarchal clustering technique which reduces distance among CHs and BS and prolongs stability period and network life, which is not discussed by authors.

Energy-efficient clustering protocols are designed by heterogeneous WSNs to prolong the network lifetime and stability period. Authors use clustering technique to reduce energy consumption in \cite{4}, CHs are elected probabilistically on the basis of ratio between residual energy of each node and average energy of network. However, we introduce hierarchal clustering technique to reduce energy consumption between CH and BS by data transmission.

\section{Existing Routing Protocols For WSNs}
Routing protocols are used to route data among sensor nodes in a network, sensed data is transmitted to CHs which further transmits aggregated data to BS. In DT routing protocol, there is no use of clustering technique to minimize energy consumption in network. Nodes sense data and directly transmit data to BS, therefore, nodes far from BS dies first. MTE is another routing approach in which those nodes which are near to sink  dies first. As a result, some part of area which is to be monitored cannot be monitored for a maximum part of network lifetime. We propose Hierarchal Stable Election Protocol (HSEP) which enhance network life and stability period than other conventional routing protocols. In this section, we discuss existing clustering based routing protocols.

\subsection{LEACH}
LEACH is self-organized, adaptive clustering protocol which uses random distribution of sensor nodes in area, to evenly distribute energy between nodes in sensor network. In LEACH, sensor nodes are organized in such a way that some of nodes become CHs which are responsible to transmit data to BS. In this process, CHs are elected on the basis of probability. Sensors nodes are elected as CHs at any given time with a certain probability which is a random number between, 0 and 1. Nodes compare this probability with a given threshold. If random number is less than threshold then sensor node becomes a CH and transmit data to BS. Otherwise, the node attach itself to any CH for communication with BS. CHs broadcast their status to other sensor nodes in network. Each sensor node join CH on basis of Received Signal Strength Indicator (RSSI). Once network is organized into clusters, each CH creates a TDMA schedule for nodes in its cluster. This allows radio components of each non CH node to be turned off at all times except during transmition time, thus, minimizing energy dissipation by individual sensors. Once CH has all aggregated data from nodes in its cluster, then CH node aggregates data and transmits compressed data to BS. Since BS is faraway in scenario which we are examining, there is high transmission energy is required however, there are only a few CHs. Therefore, a small number of nodes are affected. Being CH for a long time drains out battery of sensor nodes. To avoid this unnecessary draining of energy of single node, CHs do not remain same in all rounds. Thus, clustering seems to be an energy-efficient technique in routing protocols.

\subsection{SEP}
SEP is a routing protocol, which uses clustering based routing technique with node heterogeneity in a sense that it has fraction of advance nodes. SEP uses
a distributed fashion to select a CH in WSNs. It is heterogeneity-aware protocol and election probabilities of nodes are weighted by initial energy of each node relative to that of other nodes in a network. This enhances the time interval before death of first node; in the other words, stability period of the network. SEP is better than LEACH in evenly consuming extra energy of advanced nodes, because it gives longer stability period than LEACH which improves stable region of clustering hierarchy process using parameters of heterogeneity. In order to enhance stable region, SEP tries to maintain well and balanced energy consumption and advanced nodes become CHs more often than normal nodes. Normal nodes have initial energy equal to $E_0$, and advance nodes have $(1+a)E_0$. Where, $(a)$ is percentage of energy higher than normal nodes. In SEP, every node has some probability to become a CH. Each node selects a random number between 0 and 1, if random number become less than given threshold, $T(s)$, then that node become CH in current round to evenly distribute energy in network. $T(s)$ increases with number of rounds within each epoch and becomes equal to 1 only in last round, i.e., remaining nodes in last round become CH with probability of 1. Let us define $P_{nrm}$ as weighted election probability for normal nodes and $P_{adv}$ as weighted election probability for advanced nodes. Optimal probability of each node is divided on the basis of energy, to be elected as a CH can be calculated by using following formulas:

\begin{equation}
p_{nrm}=\frac{p_{opt}}{1+am}
\end{equation}

\begin{equation}
p_{adv}=\frac{p_{opt}}{1+am}*{(1+a)}
\end{equation}

$P_{adv}$ is probability of advance nodes to become a CH, $P_{opt}$ is optimal probability, $m$ denotes fraction of advanced nodes and $\alpha$ is an additional energy factor between advanced and normal nodes.

Now to ensure that CHs selection is done in same way as authors assume in [3], authors take another parameter into consideration, which is threshold level. Each node generates a random number between 0 and 1, if generated value is less than $T(s)$ then this node becomes a CH. All type of nodes have different formulas for calculation of threshold depending on their probabilities, which are given below:

\begin{eqnarray}
T(s_{i})=
\begin{cases}
\frac{p_{i}}{1-p_{i}(rmod\frac{1}{P_{i})}} & if\; s_{i}\epsilon G \\
0 & otherwise
\end{cases}
\end{eqnarray}

\begin{eqnarray}
T_{nrm} = \left\{ \begin{array}{rl}
 \frac{p_{nrm}}{1-p_{nrm}[r. mod \frac{1}{p_{nrm}}]} &\mbox{ if $n_{nrm} \epsilon  G'$} \\
  0 &otherwise
       \end{array} \right.
\end{eqnarray}

$G'$ denotes a set of nodes which have not become CHs in current round. $T_{nrm} $ is threshold for normal nodes to become a CH and $P_{nrm}$ is probability of normal nodes to become a CH.

\begin{eqnarray}
T_{adv} = \left\{ \begin{array}{rl}
 \frac{p_{adv}}{1-p_{adv}[r. mod \frac{1}{p_{adv}}]} &\mbox{ if $n_{adj} \epsilon  G'$} \\
  0 &\mbox{ otherwise}
       \end{array} \right.
\end{eqnarray}
$G'$ is set of nodes which have not become CHs in current round. $P_{adv}$ is probability of advance nodes to become a CH, and $T_{adv}$ is threshold for advance nodes to become a CH.

\subsection{ESEP}
ESEP is heterogeneity aware routing protocol. Energy is an important factor and must be managed in an efficient manner to prolong network life and stability period in WSNs. Authors present an easy approach which is an extension of SEP called as ESEP. In ESEP, three type of nodes are considered on the basis of their level of energy: normal, advance and intermediate nodes. Current goal of ESEP is to achieve a self configured WSN which maximizes lifetime and stability period. Major goal is to minimize communication cost and maximizing network resources to ensure correct information. Each node in a network, transmits sensed data to associated CH, whereas, CH performs data aggregation to reduce redundancy and send that data to BS. In this protocol, each sensor node chooses a random number between, 0 and 1. If this random number value is less than $T(s)$ which is given in eq. 3, then a sensor node becomes a CH in current round. In ESEP, intermediate nodes are selected in two ways authors can choose intermediate node by a relative distance of advance nodes positions to normal nodes position in network and by a threshold of energy level between advanced nodes and normal nodes.

\subsection{DEEC}
DEEC is a protocol that has been designed to deal with nodes of heterogeneous energy level in a WSN. For CH selection, DEEC uses initial and residual energy level of the nodes. Let, $n_{i}$ denotes the number of rounds to be a CH for node $s_{i}$. The protocol aims to attain $P_{opt}N$ number of CHs in network during each round. CH selection criteria in DEEC is based on energy level of the nodes. As, in homogenous network when nodes have same amount of energy during each epoch then choosing $P_{i}=P_{opt}$  assures $P_{opt}N$ CHs during each round. In heterogeneous network, the nodes with high energy are more probable to become CHs than nodes with low energy, however, the net value of CHs during each round is equal to $P_{opt}N$. $P_{i}$ is the probability for each node $s_{i}$ to become a CH, therefore, node with high energy has larger value of $P_{i}$, as compared to the $P_{opt}$. $\bar{E}(r)$ denotes average energy of network during round $R$ which is given in [4] as:

\begin{eqnarray}
\bar{E}(r)= \frac{1}{N}\sum_{i=1}^{N}E_{i}(r)
\end{eqnarray}

$p_{i}$, probability for the CH selection in DEEC is given by:

\begin{eqnarray}
p_{i}= p_{opt}[1-\frac{\bar{E}(r)-E_{i}(r)}{\bar{E}(r)}] = p_{opt}\frac{E_{i}(r)}{\bar{E}(r)}
\end{eqnarray}

In DEEC, average value of total number of CHs during each round is given in [4] as:

\begin{eqnarray}
\sum_{i=1}^{N}p_{i}= \sum_{i=1}^{N}p_{opt}\frac{E_{i}(r)}{\bar{E}(r)} = p_{opt}\sum_{i=1}^{N}\frac{E_{i}(r)}{\bar{E}(r)}= Np_{opt}
\end{eqnarray}

$G$ is set of nodes eligible to become CH at round. If node has not become CH in recent rounds then it belongs to $G'$. During each round, every node chooses a random number between 0 and 1. If the number is less than threshold, it will be become a CH else not.

As, $p_{opt}$ is reference value of average probability $p_{i}$. In homogenous networks, all nodes have same initial energy, therefore, they use $p_{opt}$ as reference energy for probability $p_{i}$. However, in heterogeneous networks, the value of  $p_{opt}$ should be different according to the initial energy of the node. In two level heterogenous network the value of $p_{opt}$ is given by:

\begin{eqnarray}
p_{adv}= \frac{p_{opt}}{1+am} , p_{nrm}= \frac{p_{opt}(1+a)}{(1+am)}
\end{eqnarray}

$p_{adv}$ and $p_{nrm}$ are used instead of $p_{opt}$ in eq. 9 for two level heterogeneous network, and is given below:

\begin{eqnarray}
p_{i}=
\begin{cases}
\frac{p_{opt}E_{i}(r)}{(1+am)\bar{E}(r)}   &   if \;s_{i}\; is\; the\; normal\; node\\
\frac{p_{opt}(1+a)E_{i}(r)}{(1+am)\bar{E}(r)}   &   if \;s_{i}\; is\; the\; advanced \;node\\
\end{cases}
\end{eqnarray}

\begin{figure}[t]
\centering
\includegraphics[height=7cm,width=8cm]{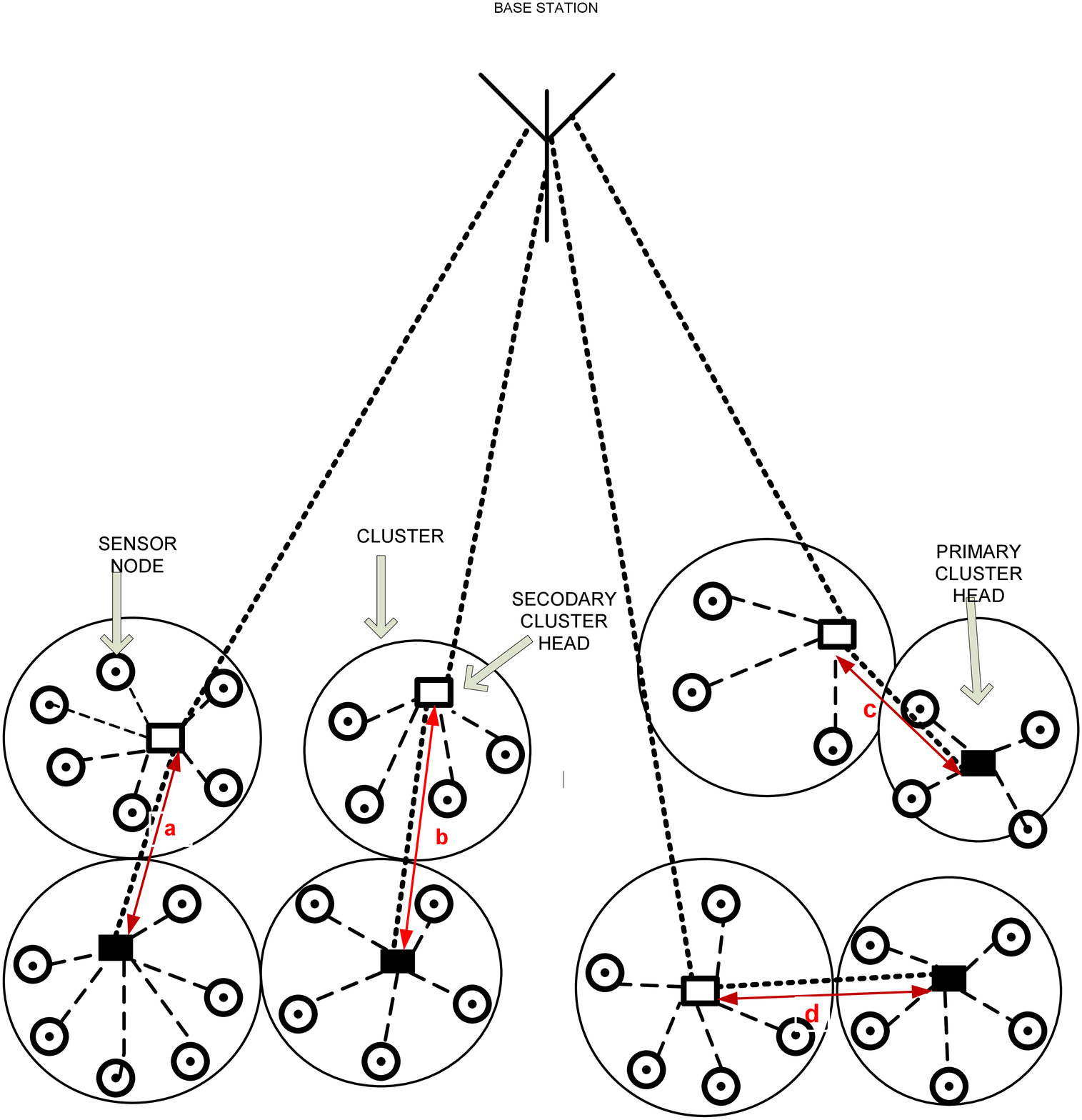}
\vspace{-0.3cm}
\caption{Network Topology}
\end{figure}

\section{HSEP: The Proposed Protocol}
HSEP is hierarchal based clustering routing protocol. As, distance between CH and BS increases it increases transmission energy, because maximum energy is consumed in process of data transmission. Our proposed protocol is aimed to reduce transmission energy between CH and BS. Therefore, we propose HSEP to minimize this transmission cost by proposing clustering hierarchy, we use two type of CHs, primary CHs and secondary CHs. HSEP is heterogeneous-aware protocol in a sense that it consists two types of nodes i.e., advance nodes and normal nodes. In this protocol these probabilities of nodes to become CHs are weighted by initial energy of a node relative to other nodes in network. This approach prolongs time interval before death of first node; in other words, stability period. Secondary CHs can be from existing primary CHs, and elect on basis of probability, ($P_h$) from those nodes which already become primary CHs and only primary CHs can take part in process of electing secondary CHs. Primary CHs check distance between each others and transmit their data to those CHs which are at minimum distance from them. However, these minimum distance CHs are secondary CHs. HSEP uses two types of nodes normal and advance nodes, advance nodes have higher probability to become CH than normal nodes. Nodes select a random number between 0 and 1, compare it with defined threshold, if random number value is less than threshold then a node become primary CH, aggregate data, send it to secondary CHs which further transmit aggregated data to BS.

Topology based two level of clustering hierarchy is used in HSEP, where, sensor nodes first sense desired data, transmit it to primary CH using TDMA slots allocated by primary CHs to their associated nodes. Primary CHS transmit their aggregated data to secondary CHS by associating with them using again TDMA slots allocated by secondary CHS, then secondary CHS further transmit aggregated data to BS. Thus minimizing transmission distance among secondary CHs and BS consumes less energy. However, whole process is define in three phases, in first phase, sensor nodes sense data according to specific requirements. This can be a temperature and motion of some body. In second phase, nodes take part to become primary CHs by comparing random number with threshold. If a node becomes primary CH, it broadcast message in network and nodes get associate with them using RSSI and send their sensed data to their CHs which we call as primary CHs. In this  phase, these primary CHs again get associate to their secondary CHs according to shortest distance between them, as shown in Fig. 1 with a, b, c and d according to this distance secondary CHs are selected, as these are the only short distances. Therefore, only these short distance, primary CHs only become secondary CH, as shown in Fig. 7, these secondary CHs aggregate data receive from primary CHs and send aggregated data to BS.

\subsection{Comparison of LEACH, SEP, ESEP, DEEC and HSEP }
To compare the efficiency of proposed protocol, we perform simulations using MATLAB. For analysis of our simulation results, we consider following performance matrices which show results for case when $m=0.1$, $\alpha=1$ and $\beta=0.3$. However, $\beta$ factor is only used in ESEP, where intermediate nodes are between normal nodes and advanced nodes. It can be easily seen from Fig. 2, stable region of HSEP is extended, as compared to LEACH, SEP, ESEP and DEEC. First node dies at $1900$ rounds, whereas, stability period of LEACH dies at $52.3\%$ less than HSEP. However, stability period of SEP is $47.3\%$ less than HSEP  and $10\%$ larger than LEACH and stability period of ESEP is $42.1\%$ less stable than HSEP, $9$ percent larger than SEP and $18$ percent larger than LEACH.

Values used for simulations are: $E_{elect}=50nJ/bit$,
$E_{DA}=5nJ/bit/message$,
$\epsilon_{fs}=10pJ/bit/m^2$,
$\epsilon_{mp}=0.0013pJ/bit/m^4$,
$E_0=0.5J$,
$K=4000$,
$P_{opt}=0.1$,
$n=100$,
$\alpha=1$,
$m=0.1$,
$Eelec=transmitter/receiver electronics energy$.
$EDA=data aggregation$,


\begin{figure}[!t]
\includegraphics[height=5cm, width =8cm]{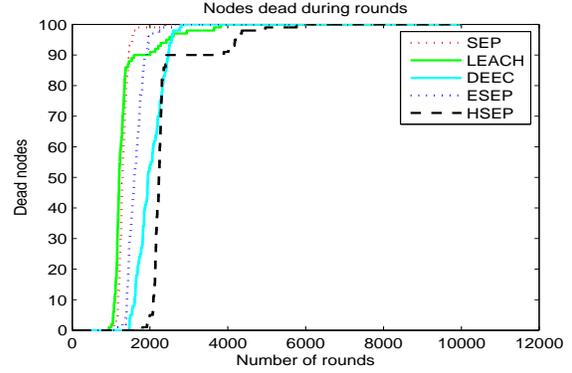}
\vspace{-0.3cm}
\caption{Comparison of HSEP with LEACH, SEP and ESEP with $\alpha=1 m=0.1$}
\end{figure}

Stability period of HSEP is $23.6\%$ larger than DEEC. However, stability period of DEEC is $24.1\%$ larger than ESEP, $31\%$ larger than SEP and $37\%$ larger than LEACH. DEEC has higher stability period than LEACH, SEP and ESEP because it uses residual energy of nodes in electing CHs, node having higher residual energy has greater chances to be a CH, therefore, stability period of DEEC is enhanced.

While ESEP, a flavor of SEP outperforms SEP and LEACH in terms of stability because three level of heterogeneity are more beneficial than two level. However, values of $\alpha$ additional energy factor between advance nodes and normal node and $\beta$ additional energy factor between advance nodes, normal nodes and intermediate nodes due to three types of nodes in ESEP it has different energy levels. If we compare ESEP and DEEC with our proposed protocol, it is observed that HSEP outperforms LEACH, SEP, DEEC and ESP in terms of stability period and also beats SEP, ESEP, LEACH and DEEC in term of network life. HSEP is out performing than others because it is hierarchal based stable election protocol in which CHs are of two level of hierarchy. In this process, once primary CHs elected then secondary CHs are elected according to defined probability and difference of distance between primary and secondary CHs. This hierarchal clustering consequently reduces transmission energy and results in large stability period and network lifetime.


In Fig. 3, there is a comparison of throughput of DEEC, HSEP, SEP, LEACH and ESEP with same parameters, as discussed above. Throughput is total number of packets send to BS from CHs in whole network life and we can see that DEEC has highest throughput. Its throughput increased in first 2500 rounds and reaches $7kbps$ and then become constant after 2500 rounds. Whereas, SEP has $1.2kbps$ throughput which is $82\%$ less than DEEC and LEACH has $1.17kbps$ throughput which is $83\%$ less than DEEC. SEP has a little bit higher throughput than LEACH because in SEP heterogeneous networks having two types of nodes which take a part in clustering, whereas, in LEACH, network is considered as homogenous. In ESEP, it has $2kbps$ throughput is achieved which is $71\%$ less than throughput of DEEC however, its throughput is higher than SEP and LEACH because of three levels of heterogeneity. Our proposed protocol has $57\%$ higher throughput than SEP, $28.5\%$ higher than ESEP and  $58.21$ higher than LEACH. Throughput of HSEP is $2.8kbps$ in 4000 rounds and become constant after 4000 rounds. Our simulation results show that HSEP outperforms ESP ,ESEP, and LEACH in throughput and DEEC out performs from all of these protocols.

\begin{figure}
\includegraphics[height=5cm, width =8cm]{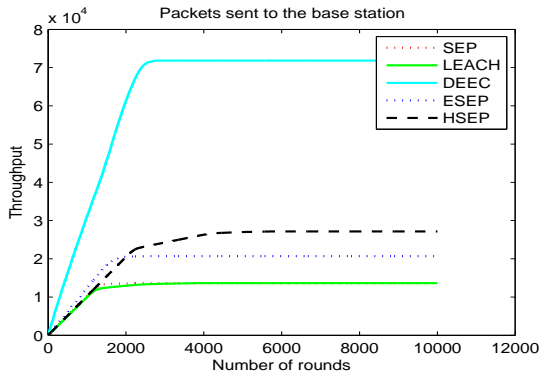}
\vspace{-0.3cm}
\caption{Comparison of HSEP with LEACH, SEP and ESEP with $\alpha=1 m=0.1$}
\end{figure}

Fig. 4 shows rate of nodes in network which are alive with number of rounds. In these results, we see that HSEP out performs DEEC, SEP, LEACH and ESEP in stability period. There is very little difference between stability period of LEACH, SEP and ESEP. However, DEEC has larger stability period than SEP, LEACH and ESEP. If we compare ESEP with SEP and LEACH, then we see that ESEP has higher stability period than SEP and LEACH because ESEP has three level of node heterogeneity, whereas SEP has two level of heterogeneity and LEACH uses homogeneous routing. Therefore, due to three level of heterogeneity, ESEP has higher stability period its first node dies at $1900$ which is $5.2\%$ more than SEP and $10\%$ more than LEACH. However, HSEP has highest network lifetime than ESEP, DEEC, LEACH and SEP. Therefore, by changing value of $\alpha$ and $m$ there is a significant improvement on network lifetime of HSEP is seen in Fig.3. Whereas, network lifetime of HSEP is $40\%$ more than ESEP , $75\%$ more than SEP network lifetime, and $61\%$ more than LEACH and $52\%$ more network life than DEEC. From our simulations, we clearly see that HSEP has largest network lifetime and stability period at $\alpha=1$ and $m=0.1$.

\begin{figure}
\includegraphics[height=5cm, width =8cm]{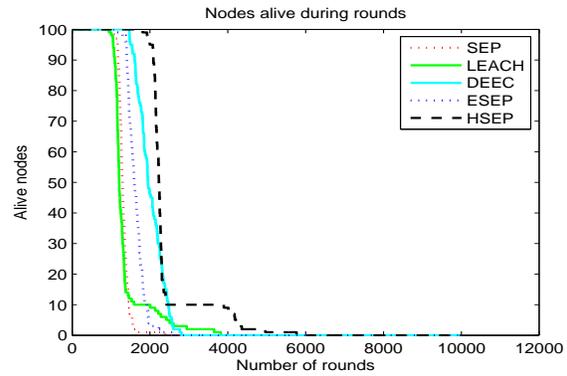}
\vspace{-0.3cm}
\caption{Comparison of HSEP with LEACH, SEP and ESEP at $\alpha=1 m=0.1$ }
\end{figure}

In Fig.5, comparison of throughput of DEEC, HSEP, SEP, LEACH and ESEP are discussed for $m=0.1$ and $\alpha=3$. We can see that DEEC has highest throughput of $14 kbps$ and is constant till end of network life. If we look at HSEP its throughput increase slowly and goes up to $5 kbps$ which is $64.2\%$ less than DEEC and then become constant after that, whereas HSEP beats SEP, LEACH and ESEP in throughput because it is hierarchal based clustered routing protocol which consumes energy more efficiently than SEP, LEACH and ESEP. Throughput of ESEP is $4kbps$ which is $71\%$ less than DEEC. However, ESEP has higher throughput than SEP and LEACH because it is heterogeneous protocol having three levels of heterogeneity. Whereas, if we talk about LEACH and SEP both have throughput of $2 kbps$ which is $85.71\%$ less than DEEC. SEP has higher throughput because it is heterogeneous protocol and have two level of heterogeneity. Whereas, LEACH is designed for homogeneous network, therefore has less throughput than SEP. It is observed from Fig. 4 and 5 that DEEC outperforms HSEP, ESEP, SEP and LEACH in throughput and stability period.
\begin{figure}
\includegraphics[height=5cm, width =8cm]{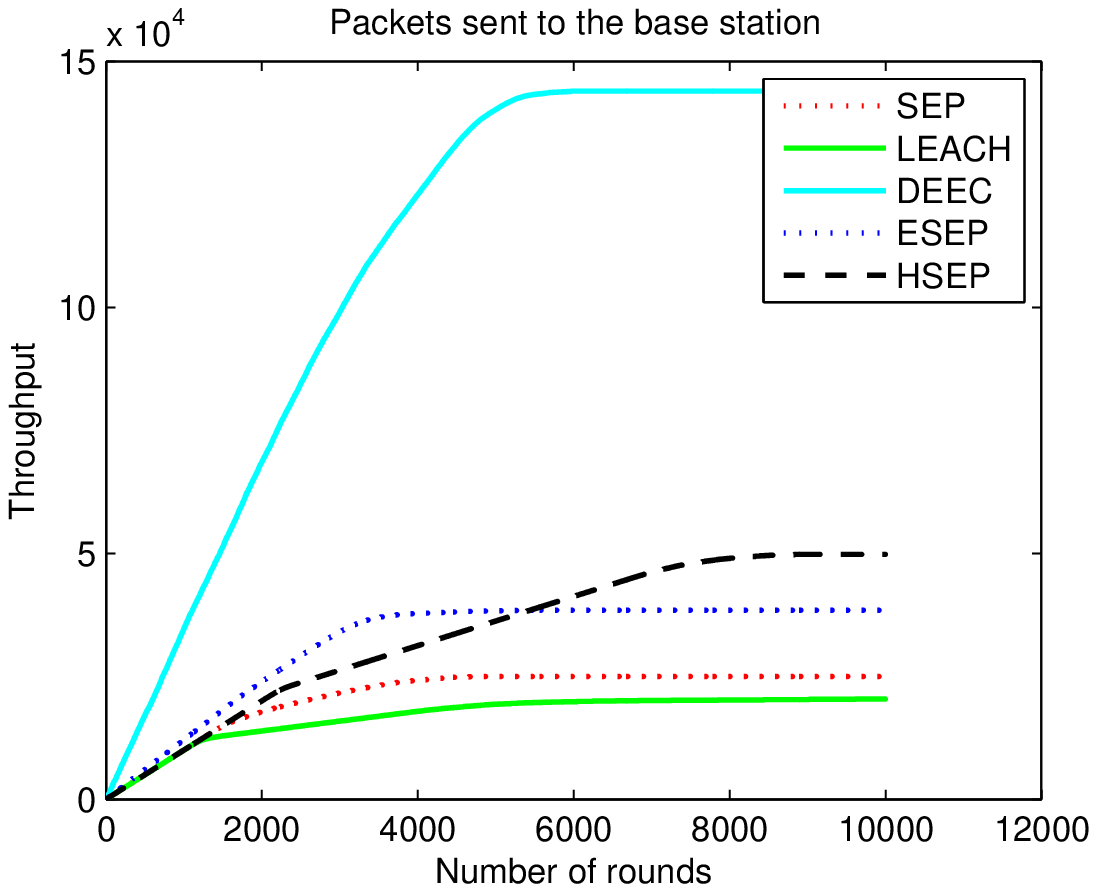}
\vspace{-0.3cm}
\caption{Comparison of HSEP with LEACH, SEP and ESEP at $\alpha=3 m=0.1$ }
\end{figure}


Characteristic parameters used in Fig. 6 shows rate of nodes in network which are alive with number of rounds. $\beta$ factor is only used in ESEP. From Fig. 6 we see that HSEP and DEEC outperforms SEP, LEACH and ESEP in stability period however, there is very low difference between first node dead round of HSEP and DEEC. HSEP has $2.6\%$ more stable than DEEC. If we talk about LEACH, SEP and ESEP, we can see that ESEP has higher stability period than LEACH and SEP because it is heterogeneous protocol. If we compare SEP with LEACH we see that SEP has higher stability period than LEACH because SEP is heterogeneous routing protocol having two level of heterogeneity and two types of nodes: advance nodes and normal nodes which take a part in clustering process. However, LEACH is homogeneous WSN protocol therefore, has same type of nodes which have equal probability to become a CH in every round and become dead early than SEP. Now, if we compare ESEP it has $5.5\%$ higher stability period than SEP and $11.1$ higher than LEACH. However, LEACH has highest network life than HSEP, ESEP and SEP. LEACH, HSEP and ESEP has $40\%$ larger than SEP and DEEC last node dies at 6000 rounds. From simulation results of Fig. 6, it is depicted that LEACH, HSEP and ESP has largest value of network lifetime and HSEP has highest stability period at $\alpha=3$ and $m=0.1$.
\begin{figure}[!ht]
\includegraphics[height=5cm, width =8cm]{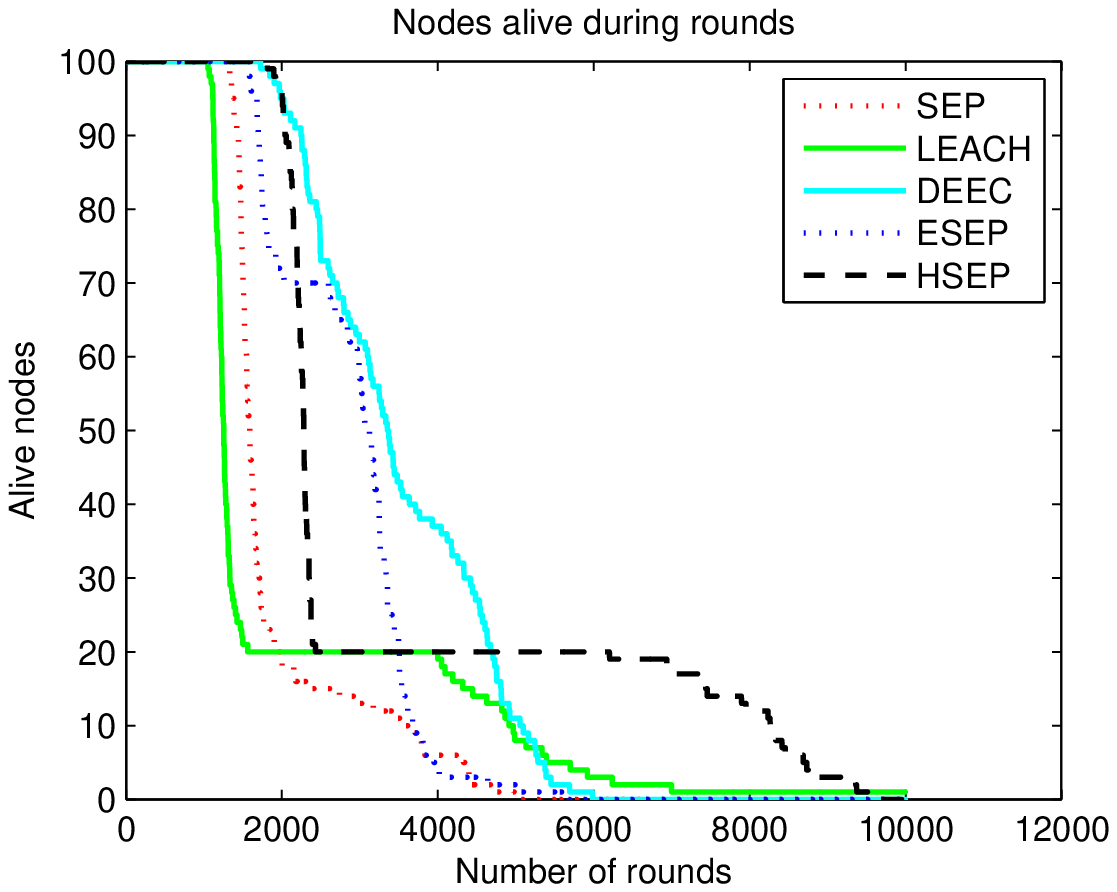}
\vspace{-0.3cm}
\caption{Comparison of HSEP with LEACH, SEP and ESEP $\alpha=3 m=0.1$}
\end{figure}
Fig. 7 shows rate of nodes in network which are going to be dead with number of rounds. It is observed that stable region of HSEP and DEEC are larger, as compared to that of LEACH, SEP and ESEP. However, there is very small difference between stable period of HSEP and DEEC, however, HSEP has $2.6\%$ more stable region than DEEC because it is hierarchal based clustering that is why energy consumption is more efficient than DEEC. ESEP has $5.5\%$ higher stability region than SEP and $11.1\%$ higher than LEACH. However, LEACH has highest network life than HSEP and DEEC. LEACH, HSEP and ESEP has $40\%$ more network lifetime than SEP and DEEC. In Fig. 7, LEACH, HSEP and ESP have largest network life and HSEP has highest stability period at $\alpha=3$ and $m=0.1$. We can see that stability period of SEP is $32\%$ less than ESEP. This period of ESEP is larger than SEP and LEACH, in which first node dies at $1500$ rounds because it is also heterogeneity awareness protocol. As three types of nodes take part in ESEP clustering, therefore, it increases stability period of network and HSEP outperforms ESEP, SEP and LEACH in stability period because of using hierarchal technique for clustering. It uses two level hierarchy in cluster formation and then transmit sensed data and efficiently utilize energy consumption in network. From our results, we can conclude that simulation HSEP has largest throughput among DEEC, SEP, LEACH and ESP, however, ESEP, LEACH and HSEP have largest network lifetime at given value of $\alpha$ and $m$.


\begin{figure}[!t]
\includegraphics[height=5cm, width =8cm]{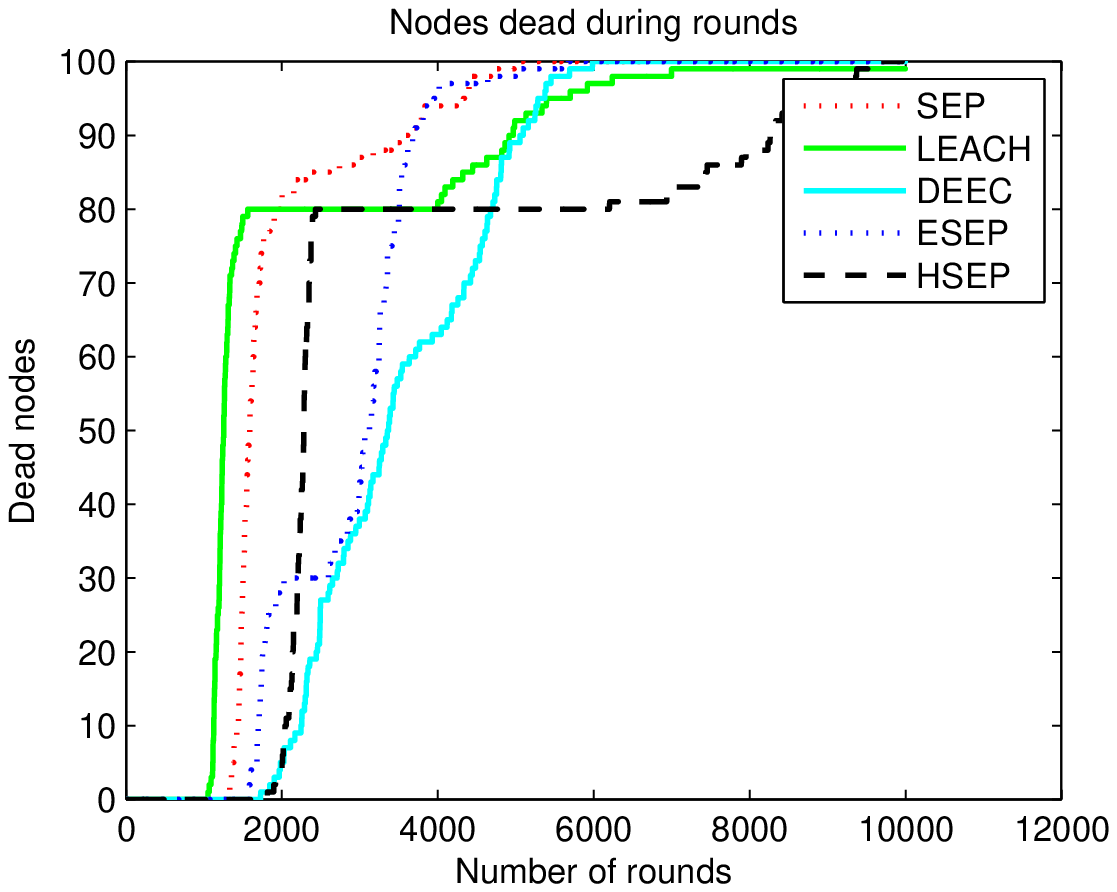}
\vspace{-0.3cm}
\caption{Comparison of HSEP with LEACH, SEP and ESEP $\alpha=3 m=0.1$}
\end{figure}

\section{Conclusion}
SEP introduces heterogeneous WSNs in which nodes have different energy levels. SEP is based on weighted election probabilities of each node to become CH according to the remaining energy. We proposed Hierarchal SEP which is also heterogenous protocol with two levels of clustering hierarchy to minimize the transmission distance between CH and sink to prolong the effective network lifetime. It is also based on weighted election probabilities of each node to become CH. We perform simulations in MATLAB to check the efficiency of our proposed protocol. Simulation results show that by simulation that HSEP always prolongs the stability period, as compared to rest of the selected protocol. Finally conclude that HSEP outperforms DEEC, SEP, ESEP and LEACH in stability period and network lifetime.


\begin{thebibliography}{1}
\bibitem{1} Heinzelman, W.R. and Chandrakasan, A. and Balakrishnan, H., ``Energy-efficient communication protocol for wireless microsensor networks'', System Sciences, Proceedings of the 33rd Annual Hawaii International Conference on, pages (10--pp), 2000.
\bibitem{2} Smaragdakis, G. and Matta, I. and Bestavros, A., ``SEP: A stable election protocol for clustered heterogeneous wireless sensor networks'', Boston University Computer Science Department, 2004.
\bibitem{3} Aderohunmu, F.A. and Deng, J.D., ``An Enhanced Stable Election Protocol (SEP) for Clustered Heterogeneous WSN''.
\bibitem{4} Qing, L. and Zhu, Q. and Wang, M., ``Design of a distributed energy-efficient clustering algorithm for heterogeneous wireless sensor networks'', Computer communications, volume (29), no. (12), pages (2230--2237), 2006.
\end{thebibliography}
\end{document}